\documentclass[10pt,journal]{IEEEtran}
\IEEEoverridecommandlockouts
\usepackage{color}

\usepackage{placeins}
\usepackage{amsthm}
\usepackage{amsmath}
\usepackage{mathtools}
\usepackage{amssymb}

\usepackage{verbatim}
\usepackage{cite}
\usepackage{url}

\usepackage{upgreek}
\usepackage{bm}
\usepackage{nicefrac}

\usepackage[named]{algo}
\usepackage[noend]{algpseudocode}
\usepackage{algorithm}

\ifCLASSOPTIONcompsoc
  \usepackage[caption=false,font=normalsize,labelfont=sf,textfont=sf]{subfig}
\else
  \usepackage[caption=false,font=footnotesize]{subfig}
\fi

\usepackage{cite}

\newtheorem{theorem}{Theorem}

\IEEEoverridecommandlockouts

\begin{document}

\title{Spectrum Sharing Radar: Coexistence via Xampling}

\author{Deborah~Cohen,
        Kumar Vijay~Mishra,
        and Yonina C.~Eldar
\thanks{The authors are with the Andrew and Erna Viterbi Faculty of Electrical Engineering, Technion - Israel Institute of Technology, Haifa, Israel, e-mail: \{debby@tx, mishra@ee, yonina@ee\}.technion.ac.il}
\thanks{This project has received funding from the European Union's Horizon 2020 research and innovation program under grant agreement No. 646804-ERC-COG-BNYQ, and from the Israel Science Foundation under Grant no. 335/14. D. C. is grateful to the Azrieli Foundation for the award of an Azrieli Fellowship. K.V.M. acknowledges partial support via Andrew and Erna Finci Viterbi Fellowship.}}


\maketitle

\begin{abstract}
This paper presents a spectrum sharing technology enabling interference-free operation of a surveillance radar and communication transmissions over a common spectrum. A cognitive radio receiver senses the spectrum using low sampling and processing rates. The radar is a cognitive system that employs a Xampling-based receiver and transmits in several narrow bands. Our main contribution is the alliance of two previous ideas, CRo and cognitive radar (CRr), and their adaptation to solve the spectrum sharing problem.
\end{abstract}

\begin{IEEEkeywords}
Spectrum sharing, spectral coexistence, cognitive radar, cognitive radio, Xampling, SSPARC
\end{IEEEkeywords}

\IEEEpeerreviewmaketitle
\vspace{-7pt}
\section{Introduction}
\label{sec:intro}
The unhindered operation of a radar that shares its spectrum with communication (``comm'', hereafter) systems has captured a great deal of attention within the operational radar community in recent years \cite{griffiths2015radar, fitz2014towards,bernhard2010final}. The interest in such \textit{spectrum sharing} radars is largely due to electromagnetic spectrum being a scarce resource and almost all services having a need for a greater access to it. With the allocation of available spectrum to newer comm technologies, the radio-frequency (RF) interference in radar bands is on the rise. Spectrum sharing radars aim to use the information from coexisting wireless and navigation services to manage this interference.

Recent research in spectrum sharing radars has focused on S and C-bands, where the spectrum has seen increasing cohabitation by Long Term-Evolution (LTE) cellular/wireless commercial comm systems. Many synergistic efforts by major agencies are underway for efficient radio spectrum utilization. The Enhancing Access to the Radio Spectrum (EARS) project by the National Science Foundation (NSF) \cite{bernhard2010final} brings together many different users for a flexible access to the electromagnetic spectrum. A significant recent development is the announcement of the Shared Spectrum Access for Radar and Comm (SSPARC) program \cite{jacyna2016ssparc, fitz2014towards} by the Defense Advanced Research Projects Agency (DARPA). This program is focused on S-band military radars and views spectrum sharing as a cooperative arrangement where the radar and comm services actively exchange information and do not ignore each other. It defines spectral \textit{coexistence} as equipping existing radar systems with spectrum sharing capabilities and \textit{spectral co-design} as developing new systems that utilize opportunistic access to the spectrum \cite{guerci2015joint}. 

A variety of system architectures have been proposed for spectrum sharing radars. Most put emphasis on optimizing the performance of either radar or comm while ignoring the performance of the other. The radar-centric architectures \cite{surender2010performance,patton2012disjoint,stinco2016spectrum} usually assume fixed interference levels from comm and design the system for high probability of detection ($P_d$). Similarly, the comm-centric systems (e.g. ``CommRad'' \cite{nartasilpa2016commrad}) attempt to improve performance metrics like the error vector magnitude (EVM) and bit/symbol error rate (BER/SER) for interference from radar \cite{metcalf2015analysis}. With the introduction of the SSPARC program, joint radar-comm performance is being investigated \cite{chiriyath2016inner,reed2016gmac,richmond2016performance}, with extensions to MIMO radar-comm \cite{li2016mimo}. In nearly all cases, real-time exchange of information between radar and comm hardware has not yet been integrated into the system architectures. Exceptions to this are automotive solutions where the same waveform is used for both target detection and comm \cite{sturm2011waveform,kumari2015investigating}. In a similar vein, our proposed method, described below, incorporates handshaking of spectral information between the two systems.

Conventional receiver processing techniques to remove RF interference in radar employ notch filters at hostile frequencies. If only a few frequencies are contaminated, then this method does not introduce exceedingly large signal distortion in radars that use wide bandwidths (e.g. FOPEN \cite{davis2011foliage}). An early work by \cite{gerlach1998thinned} suggests the use of step-frequency polyphase codes for ultrawideband radar waveforms to obtain a \textit{thinned spectrum} with nulls at interfering frequencies. Later design solutions use convex optimization of radar performance metrics for given spectral constraints (see \cite{he2012waveform} and references therein; \cite{nunn2012spectrally, rowe2014spectrally}). The objective functions in such (convex and nonconvex) optimization procedures vary, where previous studies have considered signal to noise ratio (SNR) \cite{aubry2014radar}, transmit energy in stopband \cite{patton2012disjoint}, sidelobe levels \cite{wang2011designing}, a weighted sum of suppressed band spectral energy and range sidelobes \cite{lindenfeld2004sparse,he2010waveform}, and information theoretic metrics \cite{huang2015radar,tan2016optimizing}. A recent line of research focuses on constrained quadratic program techniques to obtain a waveform that fulfills more complex spectral constraints that take into account disturbance from overlaid licensed emitters \cite{aubry2014radar, aubry2015new}. The radar is assumed to be aware of the radio environment map (REM) and optimization provides a coded transmit waveform. In all the above works, spectrum sharing is achieved by notching out the radar waveform's bandwidth causing a decrease in the range resolution.

Our spectrum sharing solution departs from this baseline. The approach we adopt follows the recently proposed Xampling (``compressed sampling") framework \cite{mishali2011xampling, SamplingBook}, a system architecture designed for sampling and processing of analog inputs at rates far below Nyquist, whose underlying structure can be modeled as a union of subspaces (UoS). The input signal belongs to a single subspace, a priori unknown, out of multiple, possibly even infinitely many, candidate subspaces. Xampling consists of two main functions: low rate analog to digital conversion (ADC), in which the input is compressed in the analog domain prior to sampling with commercial devices, and low rate digital signal processing, in which the input subspace is detected prior to digital signal processing. The resulting sparse recovery is performed using compressed sensing (CS) \cite{CSBook} techniques adapted to the analog setting. This concept has been applied to both comm \cite{mishali2010theory, mishali2011bridging, cohen2014sub, cohen2016cyclo} and radar \cite{barilan2014focusing, cohen2016towards}, among other applications.

Time-varying linear systems, which introduce both time-shifts (delays) and frequency-shifts (Doppler-shifts), such as those arising in surveillance point-target radar systems, fit nicely into the UoS model. Here, a sparse target scene is assumed, allowing to reduce the sampling rate without sacrificing delay and Doppler resolution.
The Xampling-based system is composed of an ADC which filters the received signal to predetermined frequencies before taking point-wise samples. These compressed samples, or ``Xamples", contain the information needed to recover the desired signal parameters. In \cite{barilan2014focusing,baransky2014prototype}, a multiple bandpass sampling approach was adopted that used four groups of consecutive coefficients.

Here, we capitalize on the simple observation that if only narrow spectral bands are sampled and processed by the receiver, then one can restrict the transmit signal to these. The concept of transmitting only a few subbands that the receiver processes is one way to formulate a \textit{cognitive radar} (CRr) \cite{cohen2016towards}. The delay-Doppler recovery is then performed as presented in \cite{barilan2014focusing}. The range resolution obtained through this multiband signal spectrum fragmentation is identical to that of a wideband traditional radar. Further, by concentrating all the available power in the transmitted narrow bands rather than over a wide bandwidth, the CRr increases SNR. 
In the CRr system, as detailed by \cite{cohen2016towards}, the support of subbands varies with time to allow for dynamic and flexible adaptation to the environment. Such a system also enables the radar to disguise the transmitted signal as an electronic counter measure (ECM) or cope with crowded spectrum by using a smaller interference-free portion. In this work, we focus on this latter feature. 

The CRr configuration is key to spectrum sharing since the radar transceiver can adapt its transmission to available bands, achieving coexistence with comm signals. To detect such vacant bands, a comm receiver is needed, that performs spectrum sensing over a large bandwidth. Such a task has recently received tremendous interest in the comm community, which faces a bottleneck in terms of spectrum availability. To increase the efficiency of spectrum managing, dynamic opportunistic exploitation of temporarily vacant spectral bands by secondary users has been considered, under the name of Cognitive Radio (CRo) \cite{mitola1999cognitive, haykin2005cognitive}. In this work, we use a CRo receiver to detect the occupied comm bands, so that our radar transmitter can exploit the spectral holes. One of the main challenges of spectrum sensing in the context of CRo is the sampling rate bottleneck. This issue arises since CRos typically deal with wideband signals with prohibitively high Nyquist rates. Sampling at this rate would require very sophisticated and expensive ADCs, leading to a torrent of samples. In this context, the Xampling framework provides an analog preprocessing and sub-Nyquist sampling front-end, and subsequent low rate digital recovery processing, that exploits the sparsity of the sensed signal in the frequency domain \cite{mishali2010theory}.

Here, we propose a waveform design and receiver processing solution for spectral coexistence (\textit{\`{a} la} SSPARC) composed of a comm receiver and radar transceiver implementing the Xampling concepts. The CRo comm receiver senses the spectrum from sub-Nyquist samples and provides the radar with spectral occupancy information. Equipped with this spectral map as well as a known REM detailing typical interference with respect to frequency, the CRr transmitter chooses narrow frequency subbands that minimize interference for its transmission. The delay-Doppler recovery is performed at the CRr receiver on these subbands. The combined CRo-CRr system results in spectral coexistence via the Xampling (SpeCX) framework, which optimizes the radar's performance without interfering with existing comm transmissions.

The main contribution of this work is combining two previously proposed concepts, CRo and CRr, to solve an existing practical problem, comm-radar spectrum sharing. Beyond simple combination, the CRo and CRr are adapted to the task at hand and the specific comm-radar setting. First, the CRo processing is modified to the spectrum sharing scenario of comm signal detection in the presence of radar transmissions with known support. In addition, we consider the radar transmit band selection problem conditioned to the comm detected spectrum. Finally, the CRr detection criterion, previously presented in terms of CS measures, is expressed here with respect to a radar setting. 

This paper is organized as follows. Section~\ref{sec:research} reviews spectrum sharing research. Section ~\ref{sec:prob_formulation} formulates the spectrum sharing problem and presents the comm and radar signal models. Section \ref{sec:cog_radio} introduces our CRo comm receiver that performs blind spectrum sensing. In Section \ref{sec:cog_radar}, we describe the CRr transmitted band selection and corresponding delay-Doppler recovery. Software and hardware simulations are presented in Section~\ref{sec:exp}. We conclude with a discussion on future scope in Section \ref{sec:summ}.

\section{Spectrum Sharing Across IEEE Radar Bands}
\label{sec:research}

Spectral interference to radars has drastically increased with mobile comm technology but existed long before the latter. In this section, we review some of the main spectrum sharing applications. In the VHF (30-300 MHz) and UHF (300-1000 MHz) bands, interference comes from broadcast and TV services. A common example is the FOliage PENetration (FOPEN) radar, where the receiver is conventionally designed to notch out the interfering TV/radio frequencies \cite{taylor2000ultra}. Recent introduction of the IEEE 802.11ah protocol at 900 MHz for the Internet of Things (IoT), and 802.11af in 54-790 MHz for cognitive radio technology makes VHF/UHF bands too crowded for smooth radar operation \cite{khan2016opportunistic}.

From L-band (1-2 GHz) onward, the radars begin to witness spectral intrusion from LTE. An example is the Air Route Surveillance Radar (ARSR) used by Federal Aviation Administration (FAA) sharing frequencies with WiMAX (Wireless Interoperability Microwave Access) devices \cite{wang2015atc}. Military radio services such as the Joint Tactical Information Distribution System (JTIDS) in the 969-1206 MHz band are also known to interfere with L-band radars \cite{la2013design}. However, a majority of LTE waveforms, e.g. 802.11b/g/n (2.4 GHz) WCDMA (Wide-band Code Division Multiplexing Access), WiMAX LTE, LTE GSM (Global System for Mobile comm), EDGE (Enhanced Data rates for GSM Evolution), coexist within the S-band (2-4 GHz). Therefore, most of the spectrum sharing studies are concerned with S-band radars. A recent work \cite{hessar2016spectrum} explores spectral cohabitation of Wi-Fi networks and S-band surveillance radars. LTE spectrum sharing is also being investigated for S-band shipborne air traffic control radars \cite{reed2016coexistence}. 

Spectral coexistence systems for C-band (4-8 GHz) are gradually gaining at traction due to the latest 5 GHz band allocation to 802.11a/ac VHT (Very High Throughput) wireless LAN (WLAN) technology. In particular, this is of significant concern to the Terminal Weather Doppler Radar (TDWR) network, which is co-located with the US airports \cite{labib2016coexistence}. In fact, a recent study \cite{saltikoff2015threat} identifies spectral interference threats from licensed transmitters to many other existing weather radar networks at S, C and X-bands.

At present, spectral crowding for surveillance or weather radars at frequencies higher than X-band is not under major investigation. However, in these bands, the automotive radar community has been more active in incorporating spectral cohabitation with comm services. For example, \cite{sturm2011waveform} describes the ``RadCom'' system that combines a traffic sensing K-band automotive radar with a comm link to other vehicles. At V-band, another interesting study by \cite{kumari2015investigating} shows that the 802.11ad Wi-Fi (60 GHz) Golay complementary sequence waveforms can also be used for radar remote sensing. Recently, applications of spectrum sharing in inter-vehicular comm and radar have also been proposed at W-band \cite{han2016optimal,heddebaut2010millimeter}. Furthermore, with current waveform proposals for the 5G networks, centimeter (Ka) and millimeter (V and W) wave bands are expected to become dense in the future, thus requiring innovation in shared access to the spectrum \cite{mitola20145G}. In the next section, we formulate the spectrum sharing problem, where comm and radar transmit over a common bandwidth.

\vspace{-10pt}
\section{Problem Formulation}
\label{sec:prob_formulation}
Denote the set of all frequencies of the available common spectrum by $\mathcal{F}$. The comm and radar systems occupy subsets $\mathcal{F}_C$ and $\mathcal{F}_R$ of $\mathcal{F}$, respectively. Our goal is to design the radar waveform and its support $\mathcal{F}_R$, conditional on the fact that the comm occupies frequencies $\mathcal{F}_C$. We further assume that $\mathcal{F}_C$ itself is unknown to the comm receiver, which has to first detect these frequencies. The REM is assumed known to the system as a measure of the typical spectral interference with respect to frequency. Once $\mathcal{F}_C$ is identified, the comm receiver provides a spectral map of occupied bands to the radar. Equipped with the detected spectral map and known REM, the radar waveform generator then selects the available bands with least interference for its transmission and notifies the radar receiver of its selection. The latter processes only these spectral bands using Xampling-based delay-Doppler recovery. The radar conveys the frequencies $\mathcal{F}_R$ to the comm receiver as well, so that it can ignore the radar bands while sensing the spectrum. Using our recovery methods, the radar can achieve delay-Doppler recovery performance similar to that of a radar transmitting over the entire band $\mathcal{F}$ despite using only a fraction of this bandwidth.

Our model is that of a ``friendly'' spectral coexistence where an active cooperation between radar and comm is required, as also envisaged by the SSPARC program. This is different than the spectrum sharing techniques where the two systems operate independently of each other and attempt to minimize interference in their respective spectra.
\vspace{-8pt} 
\subsection{Multiband Communication Signal}
\begin{figure}
  \centering
    \includegraphics[width=0.9\columnwidth]{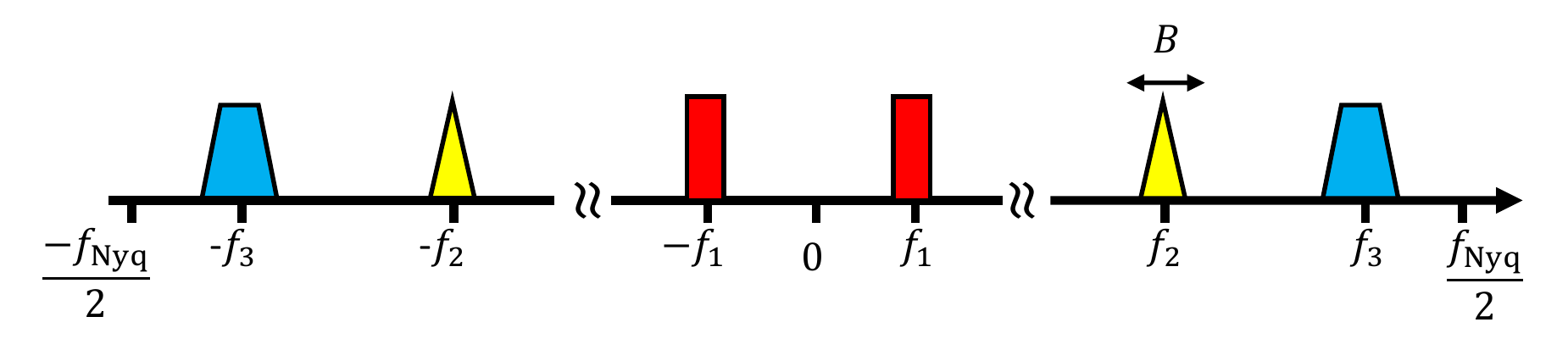}
    \vspace{-4pt}
    \caption{\scriptsize{Multiband model with $K=6$ bands. Each band does not exceed the bandwidth $B$ and is modulated by an unknown carrier frequency $|f_i| \leq f_{\text{Nyq}}/2$, for $i=1,2,3$.}\vspace{-20pt}}
    \label{fig:multiband}
\end{figure}

Let $x_{C}(t)$ be a real-valued continuous-time comm signal, supported on $\mathcal{F} = [-1/2T_{\text{Nyq}}, +1/2T_{\text{Nyq}}]$ and composed of up to $N_{\text{sig}}$ transmit waveforms such that 
\begin{equation}
\label{eq:xmodel}
x_C(t)=\sum_{i=1}^{N_{\text{sig}}} s_i(t).
\end{equation}
Formally, the Fourier transform of $x_C(t)$, defined by
\begin{equation}
X_C(f)=\lim_{T \rightarrow \infty} \frac{1}{\sqrt{T}}\int_{-T/2}^{T/2}x(t)e^{-j2\pi f t} \mathrm{d} t,
\end{equation}
is zero for every $f \notin \mathcal{F}$. We denote by $f_{\text{Nyq}} = 1/T_{\text{Nyq}}$ the Nyquist rate of $x(t)$. The waveforms, respective carrier frequencies and bandwidths are unknown. We only assume that the single-sided bandwidth $B_c^i$ for the $i$th transmission does not exceed an upper limit $B$, namely $B_c^i \leq B$ for all $1 \leq i \leq N_{\text{sig}}$.  Such sparse wideband signals belong to the so-called \textit{multiband signal model} \cite{mishali2009multicoset,mishali2010theory}. Figure~\ref{fig:multiband} illustrates the two-sided spectrum of a multiband signal with $K=2N_{\text{sig}}$ bands centered around unknown carrier frequencies $|f_i| \leq f_{\text{Nyq}}/2$. 

Let $\mathcal{F}_C \subset \mathcal{F}$ be the unknown support of $x_C(t)$, where
\begin{equation}
\mathcal{F}_C=\{f | |f-f_i|< B_c^i/2, \text{ for all } 1 \leq i \leq N_{\text{sig}}\}.
\end{equation}
The goal of the comm receiver is to retrieve $\mathcal{F}_C$, while sampling and processing $x_C(t)$ at low rates in order to reduce system cost and resources.
\vspace{-10pt}
\subsection{Pulse Doppler Radar}
Consider a standard pulse-Doppler radar that transmits a pulse train
\begin{equation}
\label{eq:uni_model}
r_{T_X}(t)= \sum_{p=0}^{P-1} h(t-p\tau), \quad 0 \leq t \leq P \tau,
\end{equation}
consisting of $P$ uniformly spaced known pulses $h(t)$. The interpulse transmit delay $\tau$ is the pulse repetition interval (PRI) (or ``fast time''); its reciprocal being the pulse repetition frequency (PRF). The entire duration of $P$ pulses in (\ref{eq:uni_model}) is known as the coherent processing interval (CPI) (or ``slow time'').

Assume that the radar target scene consists of $L$ non-fluctuating point-targets, according to the Swerling-0 target model \cite{skolnik}. The transmit signal is reflected back by the $L$ targets and these echoes are received by the radar processor. The latter aims at recovering the following information about any of the $L$ targets from the received signal:  the time delay $\tau_l$, which is linearly proportional to the range of the target from the radar; Doppler frequency $\nu_l$, proportional to the radial velocity of the target with respect to the radar; and complex amplitude $\alpha_l$, proportional to the target radar cross section, atmospheric attenuation and other propagation factors. The target locations are defined with respect to the polar coordinate system of the radar and their range and Doppler are assumed to lie in the unambiguous time-frequency region, i.e. the time delays are no longer than the PRI and Doppler frequencies are up to the PRF.
The received signal can then be written as
\begin{equation}
\label{eq:uni_rec}
r_{R_X}(t)= \sum_{p=0}^{P-1} \sum_{l=0}^{L-1} \alpha_l h(t-\tau_l - p\tau) e^{-j \nu_l p \tau}, \quad 0 \leq t \leq P\tau.
\end{equation}
It will be convenient to express $r_{R_X}(t)$ as a sum of single frames
\begin{equation}
\label{eq:frames}
r_{R_X}(t)= \sum_{p=0}^{P-1} r_{R_X}^p(t),
\end{equation}
where
\begin{equation}
\label{eq:one_frame}
r_{R_X}^p(t)= \sum_{l=0}^{L-1} \alpha_l h(t-\tau_l - p\tau) e^{-j \nu_l p \tau}, \quad p\tau \leq t \leq (p+1) \tau,
\end{equation}
is the return signal from the $p$th pulse. 

In a conventional pulse Doppler radar, the pulse $h(t)=h_{\text{Nyq}}(t)$ is a time-limited baseband function whose continuous-time Fourier transform (CTFT) is $H_{\text{Nyq}}(f)=\int_{-\infty}^{\infty} h_{\text{Nyq}}(t) e^{-j 2\pi f t} \mathrm{d}t$. It is assumed that most of the signal's energy lies within the frequencies $\pm B_h/2$, where $B_h$ denotes the effective signal bandwidth, such that the following approximation holds:
\begin{align}
H_{\text{Nyq}}(f) \approx \int\limits_{-B_h/2}^{B_h/2} h_{\text{Nyq}}(t) e^{-j 2\pi f t} \mathrm{d}t.
\end{align}
A classical radar signal processor samples each incoming frame $r_{R_X}^p(t)$ at the Nyquist rate $B_h$ to yield the digitized samples $r_{R_X}^p[n], 0 \leq n \leq N-1$, where $N=\tau B_h$. The signal enhancement process employs a matched filter for the sampled frames $r_{R_X}^p[n]$. This is then followed by Doppler processing where a $P$-point discrete Fourier transform (DFT) is performed on slow time samples. By stacking all the $N$ DFT vectors together, a delay-Doppler map is obtained for the target scene. Finally, the time delays $\tau_l$ and Doppler shifts $\nu_l$ of the targets are located on this map using a constant false-alarm rate (CFAR) detector.

The bandwidth $B_h$ of the transmitted pulses governs the range resolution of the radar. Large bandwidth is necessary to obtain high resolution, but such a spectral requirement is at odds with the coexisting comm. We, therefore, propose an alternative efficient spectral utilization method wherein the radar transmits several narrow frequency bands instead of a full-band radar signal. In particular, we propose exploiting only a fraction of the bandwidth $B_h$ for both transmission and reception of the radar signal, without degrading its range resolution. In our spectrum-sharing solution, the radar transmits a pulse $h(t)$ supported over $N_b$ disjoint frequency bands, with bandwidths $\{B_r^i\}_{i=1}^{N_b}$ centered around the respective frequencies $\{f_r^i\}_{i=1}^{N_b}$, such that $\sum_{i=1}^{N_b} B_r^i < B_h$. The number of bands $N_b$ is known to the receiver and does not change during the operation of the radar. The location and extent of the bands $B_r^i$ and $f_r^i$ are determined by the radar transmitter through an optimization procedure to identify the least contaminated bands (see Section~\ref{subsec:wf_opt}). The resulting radar transmit signal can be written as
\begin{equation}
H_R(f)
 = \left\{ \begin{array}{ll} 
    \beta_i H_{\text{Nyq}}(f), & f \in \mathcal{F}_R^i, \text{ for } 1 \leq i \leq N_b\\
    0, & \text{otherwise},
   \end{array} \right.
\end{equation}\normalsize
where $\mathcal{F}_R^i = [f_r^i -B_r^i/2, f_r^i +B_r^i/2]$ is the set of frequencies in the $i$th band. The parameters $\beta_i >1$ are chosen such that the total transmit power $P_T$ of the spectrum sharing radar waveforms remains the same as that of the conventional radar: \par\noindent\small
\begin{equation} \label{eq:pt}
\int_{-B_h/2}^{B_h/2} |H_{Nyq}(f)|^2\, \mathrm{d}f = \sum_{i=1}^{N_b}\int\limits_{\mathcal{F}_r^i} |H_R(f)|^2\, \mathrm{d}f = P_T.
\end{equation}\normalsize
In particular, if we choose $\beta_i=\beta$ for all $1 \leq i \leq N_b$ \cite{mishra2017performance}, we obtain
\begin{equation} \label{eq:beta_same}
\beta=\sqrt{\frac{\int_{-B_h/2}^{B_h/2} |H_{\text{Nyq}}(f)|^2 \mathrm{d}f}{\int\limits_{\mathcal{F}_R} |H_R(f)|^2\, \mathrm{d}f}}, 
\end{equation}
where
\begin{equation} \label{eq:omega}
\mathcal{F}_R= \bigcup_{i=1}^{N_b}\mathcal{F}_R^i.
\end{equation}
\vspace{-10pt}
\section{Cognitive Radio}
\label{sec:cog_radio}
Consider the comm signal (\ref{eq:xmodel}). When the frequency support of $x_C(t)$ is known, sampling methods such as demodulation, undersampling ADCs and interleaved ADCs \cite{SamplingBook, mishali2011bridging} can be used to reduce the sampling rate below Nyquist. When the frequency locations of the transmissions are unknown, a classic processor samples $x(t)$ at its Nyquist rate $f_{\text{Nyq}}$, which can be prohibitively high. To overcome the sampling rate bottleneck, several blind sub-Nyquist sampling and recovery schemes have been proposed that exploit the signal's structure and in particular its sparsity in the frequency domain. It has been shown \cite{mishali2009multicoset} that the minimal sampling rate for perfect blind recovery in multiband settings is twice the Landau rate \cite{landau1967density}, or twice the occupied bandwidth, namely $f_{\text{min}}=2KB=4N_{\text{sig}}B$. This rate can be significantly lower than Nyquist, by orders of magnitude.

In this work, we focus on one such technique - the modulated wideband converter (MWC) - that achieves the lower sampling rate bound. The main advantage of the MWC is that it overcomes practical issues presented by other methods, allowing its hardware implementation. We first describe the MWC sub-Nyquist sampling scheme and then turn to signal recovery from low rate samples. We begin with a scenario where the radar is silent so that the signal sensed by the comm receiver is $x_C(t)$ and then extend our approach to include spectrum sensing in the presence of a known radar signal.
\vspace{-8pt} 
\subsection{Sub-Nyquist Sampling}
\begin{figure}
	\centering
		\includegraphics[width=1\columnwidth]{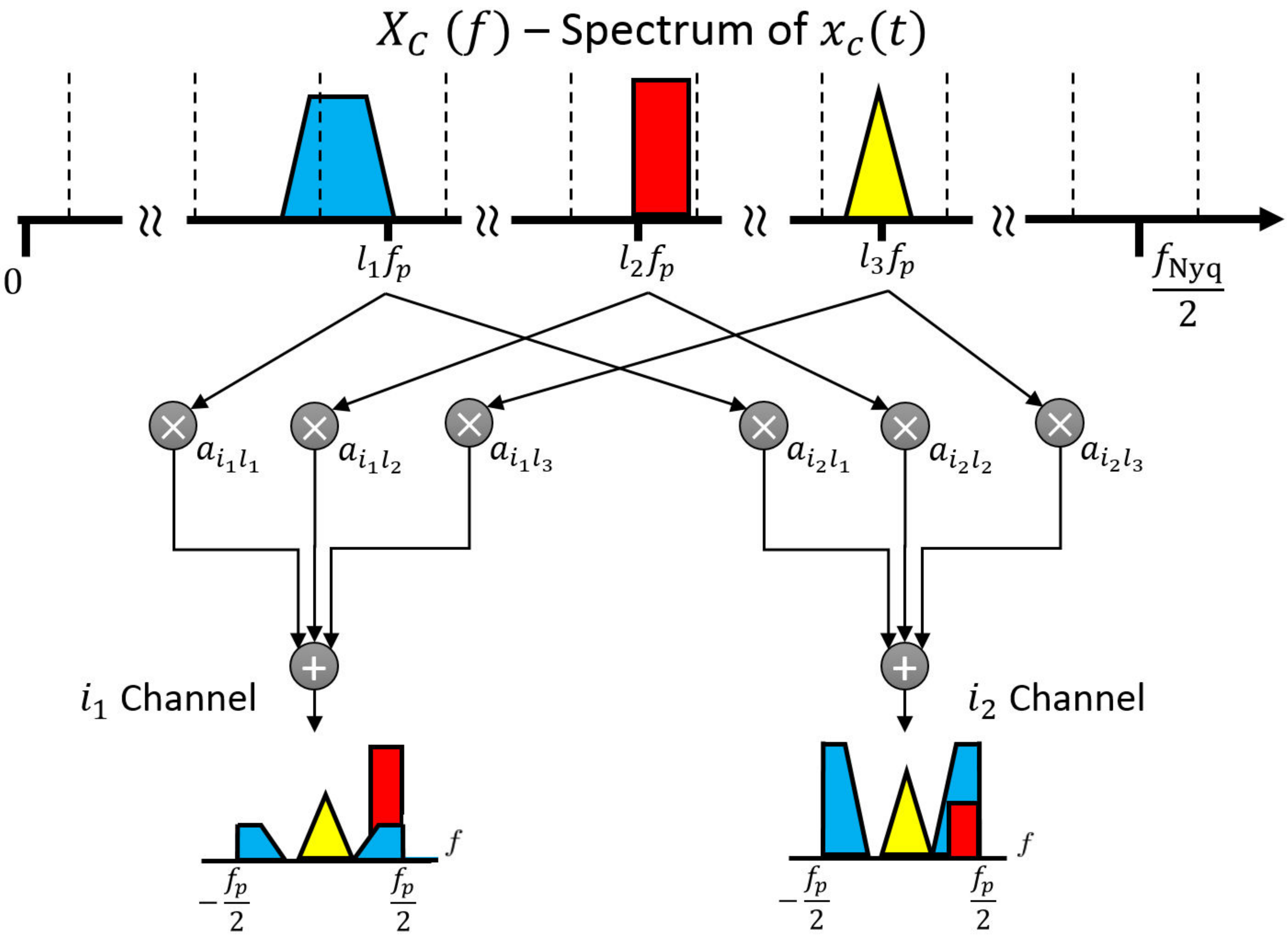}
        \vspace{-10pt}
		\caption{\scriptsize{Spectrum slices of the input signal $\mathbf{x}_C(f)$ multiplied by the coefficients $a_{il}$ of the sensing matrix $\mathbf{A}$, resulting in the measurements $z_i(f)$ for the $i$th channel.}}
        \vspace{-10pt}
		\label{fig:zAx}
\end{figure}

The MWC \cite{mishali2010theory} is composed of $M$ parallel channels. In each channel, an analog mixing front-end, where $x_C(t)$ is multiplied by a mixing function $p_i(t)$, aliases the spectrum, such that each band appears in baseband. The mixing functions $p_i(t)$ are periodic with period $T_p$ such that $f_p=1/T_p \ge B$ and have thus the following Fourier expansion:
\begin{equation}
p_i(t) =\sum_{l=-\infty}^{\infty} c_{il} e^{j\frac{2\pi}{T_p} lt}.
\end{equation}
In each channel, the signal goes through a lowpass filter (LPF) with cut-off frequency $f_s/2$ and is sampled at the rate $f_s \ge f_p $, resulting in the samples $z_i[n]$. Define 
\begin{equation}
\label{eq:Ndef}
N=2\left\lceil \frac{f_{\text{Nyq}}+f_s}{2f_p} \right\rceil,
\end{equation}
and $\mathcal{F}_s=[-f_s/2,f_s/2]$.
Following the calculations in \cite{mishali2010theory}, the relation between the known discrete time Fourier transform (DTFT) of the samples $z_i[n]$ and the unknown $X_C(f)$ is given by
\begin{equation} \label{eq:mwc}
\mathbf{z}(f)=\mathbf{A}\mathbf{x}_C(f), \qquad f \in \mathcal{F}_s,
\end{equation}
where $\mathbf{z}(f)$ is a vector of length $N$ with $i$th element $z_i(f)=Z_i(e^{j2\pi fT_s})$ and the unknown vector $\mathbf{x}_C(f)$ is given by
\begin{equation}
{\mathbf{x}_C}_i(f)=X_C(f+(i-\lceil N/2 \rceil)f_p), \quad f \in \mathcal{F}_s,
\end{equation}
for $1 \leq i \leq N$.
This relation is illustrated in Fig.~\ref{fig:zAx}. The $M \times N$ matrix $\mathbf{A}$ contains the known coefficients $c_{il}$ such that
\begin{equation}
\mathbf{A}_{il} = c_{i,-l}=c^*_{il}.
\end{equation}

The minimal number of channels to recover the $K$-sparse vector $\mathbf{x}_C(f)$, for $f \in \mathcal{F}_s$, dictated by CS results \cite{CSBook}, is $M \geq 2K$ with $f_s\geq B$ per channel. The overall sampling rate, given by
\begin{equation}
f_{\text{tot}}=Mf_s=\frac{M}{N}f_{\text{Nyq}},
\end{equation}
with $M<N$, can thus be as low as $f_{\text{min}}=2KB \ll f_{\text{Nyq}}$.

The number of branches $M$ dictates the total number of hardware devices and thus governs the level of hardware complexity. Reducing the number of channels is thus a crucial challenge for the practical implementation of a CRo receiver. The MWC architecture presents an interesting flexibility property that permits trading channels for sampling rate, allowing to drastically reduce the number of channels.  
\begin{figure*}[!t]
	\centering
		\includegraphics[width=\textwidth]{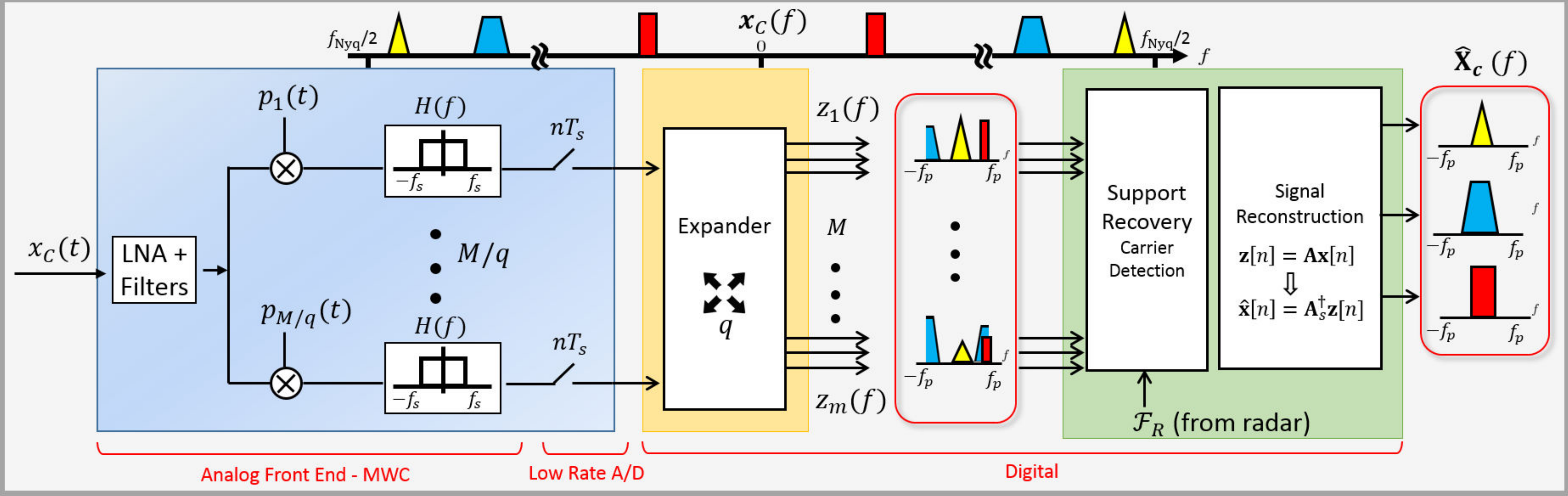}
		\caption{\scriptsize{Schematic implementation of the MWC analog sampling front-end and digital signal recovery from low rate samples. The CRo inputs are the comm signal $x_C(t)$ and radar support $\mathcal{F}_R$. The comm support output $\mathcal{F}_C$ is shared with the radar transmitter.}\vspace{-10pt}}
		\label{fig:HighLevel}
\end{figure*}
Consider a configuration where $f_s=qf_p$, with odd $q$. In this case, the $i$th physical channel provides $q$ equations over $\mathcal{F}_p=[-f_p/2,f_p/2]$. Conceptually, $M$ physical channels sampling at rate $f_s=qf_p$ are then equivalent to $Mq$ channels sampling at $f_s=f_p$. The output of each of the $M$ physical channels is digitally demodulated and filtered to produce samples that would result from $Mq$ equivalent virtual branches. This happens in the so-called expander module, directly after the sampling stage. The number of channels is thus reduced at the expense of higher sampling rate $f_s$ in each channel and additional digital processing. At its brink, this strategy allows to collapse a system with $M$ channels to a single branch with sampling rate $f_s=Mf_p$ (further details can be found in \cite{SamplingBook, mishali2010theory,mishali2011hardware}).

The MWC analog mixing front-end, shown in Fig.~\ref{fig:HighLevel}, results in folding the spectrum to baseband with different weights for each frequency interval. The goal is now to recover $x_C(t)$, or alternatively $\mathbf{x}_C(f)$, from the low rate samples. In the next section, we provide a reconstruction algorithm that achieves the minimal rate of $2KB$.
\vspace{-10pt} 
\subsection{Signal Recovery}
It is interesting to note that (\ref{eq:mwc}), which is written in the frequency domain, is valid in the time domain as well. We can therefore reconstruct $\mathbf{x}_C(f)$ in the frequency domain, or alternatively, recover $\mathbf{x}_C[n]$ in the time domain using
\begin{equation}
\mathbf{z}[n] = \mathbf{A} \mathbf{x}_C[n].
\label{eq:time}
\end{equation}

The systems (\ref{eq:mwc}) and (\ref{eq:time}) are underdetermined due to the sub-Nyquist setup and known as infinite measurement vectors (IMV) in the CS literature \cite{CSBook, SamplingBook}. With respect to these two properties, the digital reconstruction algorithm encompasses the following three stages \cite{SamplingBook, mishali2009multicoset} that we explain in more detail below:
\begin{enumerate}
\item The continuous-to-finite (CTF) block constructs a finite frame (or basis) from the samples.
\item The support recovery formulates an optimization problem whose solution's support is identical to the support $S_C$ of $\mathbf{x}_C[n]$, that is the active slices.
\item The signal can then be digitally recovered by reducing (\ref{eq:time}) to the support of $\mathbf{x}_C[n]$.
\end{enumerate}

The recovery of $\mathbf{x}_C[n]$ for every $n$ or $\mathbf{x}_C(f)$ for each $f$ independently is inefficient and not robust to noise. Instead, the support recovery paradigm from \cite{mishali2009multicoset} exploits the fact that the bands occupy continuous spectral intervals so that $\mathbf{x}_C(f)$ are jointly sparse for $f \in \mathcal{F}_p$, that is they have the same spectral support $S_C$. The CTF block \cite{mishali2009multicoset} then produces a finite system of equations, called multiple measurement vectors (MMV) from the infinite number of linear systems (\ref{eq:mwc}) or (\ref{eq:time}).

From (\ref{eq:mwc}) or (\ref{eq:time}), we have
\begin{equation}
\mathbf{Q = \Phi Z}_C \mathbf{\Phi}^H
\end{equation}
where
\begin{equation} \label{eq:q_ctf}
\mathbf{Q}= \int_{f \in \mathcal{F}_p} \mathbf{z}(f) \mathbf{z}^H(f) \mathrm{d}f = \sum_{n=\infty}^{\infty} \mathbf{z}[n] \mathbf{z}^H[n]
\end{equation}
is a $M \times M$ matrix and
\begin{equation}
\mathbf{Z}_C= \int_{f \in \mathcal{F}_p} \mathbf{x}_C(f) \mathbf{x}_C^H(f) \mathrm{d}f = \sum_{n=-\infty}^{\infty} \mathbf{x}_C[n] \mathbf{x}_C^H[n]
\end{equation}
is a $N \times N$ matrix. 
The matrix $\bf Q$ is then decomposed to a frame $\bf V$ such that $\mathbf{Q=VV}^H$. Clearly, there are many possible ways to select $\bf V$. One possibility is to construct it by performing an eigendecomposition of $\bf Q$ and choosing $\bf V$ as the matrix of eigenvectors corresponding to the non zero eigenvalues. The finite dimensional MMV system
\begin{equation} \label{eq:CTF}
\mathbf{V}=\mathbf{A}\mathbf{U}_C,
\end{equation}
is then solved for the sparsest matrix $\mathbf{U}_C$ with minimal number of non-identically zero rows using CS techniques \cite{CSBook, SamplingBook}. The key observation of this strategy is that the support of the unique sparsest solution of (\ref{eq:CTF}) is the same as the support of our original set of equations (\ref{eq:mwc}) \cite{mishali2009multicoset}. Recovering $\mathbf{U}_C$ from (\ref{eq:CTF}) can be performed using any MMV CS algorithm such as simultaneous orthogonal matching pursuit (SOMP) and simultaneous iterative hard thresholding (SIHT) \cite{CSBook}.

Note that $\mathbf{x}_C(f)$ is $K$-sparse for each specific frequency $f \in \mathcal{F}_p$, whereas $\mathbf{x}_C[n]$ is $2K$-sparse since each transmission can split between two bins, as shown in Fig.~\ref{fig:zAx} for the blue trapeze. After combining the frequencies, the matrix $\mathbf{U}_C$ is $2K$-sparse (at most) as well. Therefore, the above algorithm, referred to as SBR4 in \cite{mishali2009multicoset} requires a minimal sampling rate of $2f_{\text{min}}$. In order to achieve the minimal rate $f_{\text{min}}$, the SBR2 algorithm regains the factor of two in the sampling rate at the expense of increased complexity \cite{mishali2009multicoset}. In a nutshell, SBR2 is a recursive algorithm that alternates between the CTF described above and a bi-section process. The bi-section splits the original frequency interval into two equal width intervals on which the CTF is applied, until the level of sparsity of $\mathbf{U}_C$ is less or equal to $K$. As opposed to SBR4 which can be performed both in time and frequency, SBR2 can obviously be performed only in the frequency domain. We refer the reader to \cite{mishali2009multicoset} for more details. 

Once the support $S_C$ is known, the slices of $x_C(t)$ are recovered either in the frequency or time domain by reducing the system of equations (\ref{eq:mwc}) or (\ref{eq:time}), respectively, to $S_C$. In the time domain, we have 
\begin{eqnarray} \label{eq:recs}
\mathbf{\hat{x}}^{S_C}_C[n] &=& \mathbf{A}_{S_C}^{\dagger} \mathbf{z}[n], \\
\mathbf{\hat{x}}_{C_i}[n] &=& 0, \quad \forall i \notin {S_C}. \nonumber
\end{eqnarray}
Here, $\mathbf{x}^{S_C}_C[n]$ denotes the vector $\mathbf{x}_C[n]$ reduced to its support, $\mathbf{A}_{S_C}$ is composed of the columns of $\bf A$ indexed by $S_C$ and $\dagger$ is the Moore-Penrose pseudo-inverse. The occupied comm support is then given by
\begin{equation}  \label{eq:Fc}
\mathcal{F}_C=\{ f | |f-(i+\left\lceil N/2 \right\rceil)f_p| \leq \frac{f_p}{2}, \text{ for all } i \in S_C \}.
\end{equation}
A finer support can be estimated by performing energy detection on the recovered bands $\mathbf{\hat{x}}^{S_C}_C(f)$ for $f \in \mathcal{F}_p$.
Last, if needed, the Nyquist rate samples $x[n]=x(nT_{\text{Nyq}})$ can be reconstructed by summing the modulated and interpolated sequences $\mathbf{x}_C[n]$ to the Nyquist rate, as
\begin{equation}
x[n]=\sum_{i \in S_c} (\mathbf{\hat{x}}_{C_i}[n] * h_I[n])e^{j 2 \pi f_p nT_{\text{Nyq}}},
\end{equation}
where $h_I[n]$ is the digital interpolation filter.
The MWC sampling and recovery processes are illustrated in Fig.~\ref{fig:HighLevel}.

\vspace{-10pt} 
\subsection{Communication Signal Recovery in the Presence of Radar Transmission}
In the previous section, we considered the scenario where the radar is silent and only the comm signal $x_C(t)$ is received. Here, we treat a more general setting in which the received signal is given by
\begin{equation}
x(t)=x_C(t)+x_R(t),
\end{equation}
where $x_R(t)=r_{T_X}(t)+r_{R_X}(t)$ is the radar signal sensed by the comm receiver, composed of the transmitted and received radar signals defined in (\ref{eq:uni_model}) and (\ref{eq:uni_rec}), respectively.
Following the derivations from the previous section, we can write the sub-Nyquist samples in the Fourier domain as
\begin{equation} \label{eq:mwc2}
\mathbf{z}(f)=\mathbf{A}(\mathbf{x}_C(f)+\mathbf{x}_R(f)), \qquad f \in \mathcal{F}_s,
\end{equation}
where
\begin{equation}
{\mathbf{x}_R}_i(f)=X_R(f+(i-\lceil N/2 \rceil)f_p), \quad 1 \leq i \leq N, f \in \mathcal{F}_s.
\end{equation}
The equation solved by the CTF then becomes
\begin{equation} \label{eq:CTF2}
\mathbf{V}=\mathbf{A}(\mathbf{U}_C + \mathbf{U}_R).
\end{equation}

The frequency support $\mathcal{F}_R$ of $x_R(t)$, given by (\ref{eq:omega}), is known at the comm receiver. From $\mathcal{F}_R$, we derive the support $S_R$ of the radar slices $\mathbf{x}_R(f)$, which is identical to the support of $\mathbf{U}_R$, such that
\begin{equation} \label{eq:sr}
S_R= \left\{ n \left|  \left|n - \frac{f_R^i}{f_p}-\left\lceil N/2 \right\rceil \right| \right. <\frac{f_s+B_R^i}{2f_p} \right\}
\end{equation}
for $1 \leq i \leq N_b$.
Our goal can then be stated as recovering the support of $\mathbf{U}_C$ from $\bf V$, given the known support $S_R$ of $\mathbf{U}_R$. This can be formulated as a sparse recovery with partial support knowledge, studied under the framework of modified CS \cite{vaswani2010modified,vaswani2016recursive}. From \cite{vaswani2010modified}, the minimal number of channels required for the exact reconstruction of the $K$-sparse matrix $\mathbf{U}_C$ is $M \geq 2K + |S_R|$. Note that \cite{vaswani2010modified} considers a single measurement vector (SMV) and we extend this result to MMV.

The modified-CS idea has been used to adapt CS recovery algorithms to exploit the partial known support a priori information. In particular, greedy algorithms, such as OMP and IHT have been modified to OMP with partial known support (OMP-PKS) \cite{stankovi2009compressive} and IHT-PKS \cite{carrillo2010iterative}, respectively. In OMP-PKS, instead of starting with an initial empty support set, one starts with $S_R$ as being the initial support set. The remainder of the algorithm is then identical to OMP. In each iteration of IHT-PKS, the estimator over the known support is kept and the thresholding is performed only over the complementary support.
Algorithm \ref{algo:comm} summarizes the resulting sub-Nyquist comm signal recovery in the presence of radar transmission, using OMP-PKS for support recovery. 

\begin{algorithm}[!ht]
\caption{Cognitive Radio Spectrum Sensing}\label{algo:comm} 
	\begin{algorithmic}[1]
		\qinput Observation vector $\mathbf{z}(f)$, $f \in \mathcal{F}_s$, radar support $\mathcal{F}_R$
		\qoutput Comm signal support $\mathcal{F}_C$ and slices estimate $\mathbf{\hat{x}}_C[n]$
		\State Compute the support $S_R$ as in (\ref{eq:sr})
		\State Compute $\bf Q$ from (\ref{eq:q_ctf}) and extract a frame $\bf V$ such that $\mathbf{Q}=\mathbf{V}\mathbf{V}^H$ using eigendecomposition
        \State Compute the estimate $$\mathbf{\hat{U}}_1^{S_R} = \mathbf{A}_{S_R}^{\dagger}\mathbf{V}, \quad
\mathbf{\hat{U}}_{1_i} = \mathbf{0}, \quad \forall i \notin {S_R}$$
		\State Compute the residual $$ \mathbf{V}_1=\mathbf{V}-\mathbf{A}_{S_R}\mathbf{\hat{U}}_1$$
        \State Find the total signal support $S_R \bigcup S_C$ using OMP from the $2$nd iteration with sampling matrix $\bf A$, residual $\mathbf{V}_1$ and support $S_R$
		\State Find the comm (and radar) slices estimate from 
\begin{eqnarray}
\mathbf{\hat{x}}^{S_C \bigcup S_R}[n] &=& \mathbf{A}_{S_C \bigcup S_R}^{\dagger} \mathbf{z}[n], \nonumber \\
\mathbf{\hat{x}}_{i}[n] &=& 0, \quad \forall i \notin {S_C \bigcup S_R}. \nonumber
\end{eqnarray}              
		\State Compute the comm signal support $\mathcal{F}_C$ from (\ref{eq:Fc})
	\end{algorithmic}
\end{algorithm}
\vspace{-10pt}
\section{Cognitive Radar}
\label{sec:cog_radar}
Once the set $\mathcal{F}_C$ is estimated, the objective of the radar is to identify an appropriate transmit frequency set $\mathcal{F}_R \subset \mathcal{F}\setminus\mathcal{F}_C$ such that the radar's probability of detection $P_d$ is maximized. For a fixed probability of false alarm $P_{\text{fa}}$ the $P_d$ increases with higher signal to interference and noise ratio (SINR) \cite{kay1998fundamentals}. Hence, the frequency selection process can, alternatively, choose to maximize the SINR or minimize the spectral power in the undesired parts of the spectrum. At the receiver of this spectrum sharing radar, we employ the sub-Nyquist approach described in \cite{barilan2014focusing}, where the delay-Doppler map is recovered from the subset of Fourier coefficients defined by $\mathcal{F}_R$.
\vspace{-10pt} 
\subsection{Optimal Radar Transmit Bands}
\label{subsec:wf_opt}
The REM is assumed to be known to the radar transmitter in the form of typical interfering energy levels with respect to frequency bands, represented by a vector $\mathbf{y} \in \mathbb{R}^q$, where $q$ is the number of frequency bands with bandwidth $b_y \triangleq |\mathcal{F}|/q$.
In addition, the information available from the CRo indicates that the radar waveform must avoid all the frequencies in the set $\mathcal{F}_C$. Therefore, we further set $\bf y$ to be equal to $\infty$ in the bands that coincide with $\mathcal{F}_C$.
The goal is now to select subbands from the set $\mathcal{F}\setminus\mathcal{F}_C$ with minimal interference energy. We thus seek a block-sparse frequency vector $\mathbf{w} \in \mathbb{R}^p$ with unknown block lengths, where $p$ is the number of discretized frequencies, and whose support provides frequency bands with low interference for the radar transmission. Each entry of $\bf w$ represents a frequency subband of bandwidth $b_w \triangleq |\mathcal{F}|/p$.

To this end, we use the structured sparsity framework from \cite{huang2011learning}  that extends standard sparsity regularization to structured sparsity. We adopt the one-dimensional graph sparsity structure whose nodes are the ordered entries of $\mathbf{w}$, so that neighbor nodes are indexed by adjacent frequency bands. The graph dimension is therefore the frequency and its size is $p$. Block sparsity may be enforced by allowing the graph to contain connected regions. In contrast to traditional block sparsity approaches [30], this formulation does not require \textit{a priori} knowledge on the location of the non-zero blocks. This is achieved by replacing the traditional sparse recovery $\ell_0$ constraint by a more general term $c(\mathbf{w})$, referred to as the coding complexity, such that
\begin{equation} \label{eq:c_def} 
c(\mathbf{w})= \min \{c(F) | \text{supp}(\mathbf{w}) \subset F \},
\end{equation} 
where $F \subset \{1, \dots, p\}$ is a sparse subset of the index set of the coefficients of $\bf w$. In particular, for graph sparsity, the choice of $c(F)$ is simply \par\noindent\small
\begin{equation} \label{eq:cF_def} 
c(F)=g\log p +|F|,
\end{equation} \normalsize
where $g$ is the number of connected regions, or blocks, of $F$. This coding complexity favors blocks within the graph. 

The resulting optimization problem for finding the block-sparse frequency vector $\mathbf{w}$ can then be expressed as



\begin{equation}
	\label{eq:band_select1}
	 \min \; ||\mathbf{y}_{\text{inv}}-\mathbf{D}\mathbf{w}||_2^2 + \lambda c(\mathbf{w}),
\end{equation}
where $\lambda$ is a regularization parameter and $c(\mathbf{w})$ is defined in (\ref{eq:c_def}) with $c(F)$ in (\ref{eq:cF_def}). Here, $\mathbf{y}_{\text{inv}}$ contains element-wise reciprocals of $\bf y$, namely $(\mathbf{y}_{\text{inv}})_i = 1/\mathbf{y}_i$, so that small values in $\mathbf{y}_{\text{inv}}$ induce corresponding zero blocks in $\mathbf{w}$, and $\bf D$ is a $q \times p$ matrix that maps each discrete frequency in $\bf w$ to the corresponding band in $\mathbf{y}_{\text{inv}}$. That is, the $(i,j)$th entry of $\mathbf{D}$ is equal to $1$ if the $j$th frequency in $\bf w$ belongs to the $i$th band in $\bf y$; otherwise, it is equal to $0$. If we choose $p=q$, then $\bf D=I$ is the $q \times q$ identity matrix.

Problem (\ref{eq:band_select1}) can be solved using a structured greedy algorithm, structured OMP (StructOMP), presented in \cite{huang2011learning} and adapted to our setting in Algorithm \ref{algo:sel}. In \cite{huang2011learning}, the algorithm proceeds by greedily adding blocks one at a time to reduce the loss, scaled by the cost of the added block. Here, we consider single element blocks for simplicity but larger blocks can be considered to increase the algorithm's effectiveness. 
In the original StructOMP \cite{huang2011learning}, the stopping criterion is based on additional \textit{a priori} information on the overall sparsity and number of non-zero blocks. We adopt an alternative stopping criterion, based only on the number of blocks, which is known to be equal to $N_b$ in our problem. 
This leads to $N_b$ bands in $\mathcal{F}_R$ as dictated by the hardware constraints.
In the above, additional requirements of transmit power constraints, range sidelobe levels, and minimum separation between the bands may also be imposed and, if needed, alternative structured greedy algorithms that require very little \textit{a priori} knowledge of block sparsity can be used \cite{yu2012bayesian, peleg2012exploiting}.  
Once the support $\mathcal{F}_R$ is identified, a suitable waveform code may be designed using optimization procedures described by e.g. \cite{aubry2014radar,he2010waveform}.

\begin{algorithm}[!ht]
\caption{Cognitive Radar Band Selection}\label{algo:sel} 
	\begin{algorithmic}[1]
		\qinput REM vector $\mathbf{y}$ and subbands bandwidth $b_y$, shared support $\mathcal{F}$, comm support $\mathcal{F}_C$, mapping matrix $\bf D$, number of discretized frequencies $p$, number of bands $N_b$
		\qoutput Block sparse vector $\bf w$, radar support $\mathcal{F}_R$
		\State Set $\mathbf{y}_i = \infty$, for all $1 \leq i \leq q$ such that $[ib_y-|\mathcal{F}|/2, (i+1)b_y-|\mathcal{F}|/2[\bigcap \mathcal{F}_C \neq \emptyset$ and compute $(\mathbf{y}_{\text{inv}})_i=1/\mathbf{y}_i$
		\State Initialization $F_0 = \emptyset$, $\bf w=0$, $t=1$
        \State Find the index $\lambda_t$ so that $\lambda_t = \arg \max \phi(i)$, where
        $$ \phi(i)=\frac{||\mathbf{P}_i(\mathbf{D}\mathbf{\hat{w}}_{t-1}-\mathbf{y}_{\text{inv}})||_2^2}{c(i \bigcup F_{t-1})-c(F_{t-1})}$$ with $\mathbf{P}_i=\mathbf{D}_i(\mathbf{D}_i^T\mathbf{D}_i)^{\dagger}\mathbf{D}_i^T$
        \State Augment index set $F_t = \lambda_t \bigcup F_{t-1}$
         \State Find the new estimate $$\mathbf{\hat{w}}_{t|F_t}=\mathbf{D}_{F_t}^{\dagger} \mathbf{y}_{\text{inv}}, \quad \hat{\mathbf{w}}_{t|F_t^C}= \mathbf{0}$$
         \State If the number of blocks, or connected regions, $g(\mathbf{w}) >N_b$, go to step 7. Otherwise, return to step 3
         \State Remove the last index $\lambda_t$ so that $F_t = F_{t-1}$ and $\mathbf{\hat{w}}_t= \mathbf{\hat{w}}_{t-1}$ 
         \State Compute the radar support $$\mathcal{F}_R= \bigcup_{j \in F_t} [jb_w-|\mathcal{F}|/2, (j+1)b_w-|\mathcal{F}|/2[$$
	\end{algorithmic}
\end{algorithm}



\vspace{-6pt} 
\subsection{Delay-Doppler Recovery}
\label{subsec:delaydoprec}
\begin{figure}[!t]
	\centering
	\includegraphics[width=0.32\textwidth]{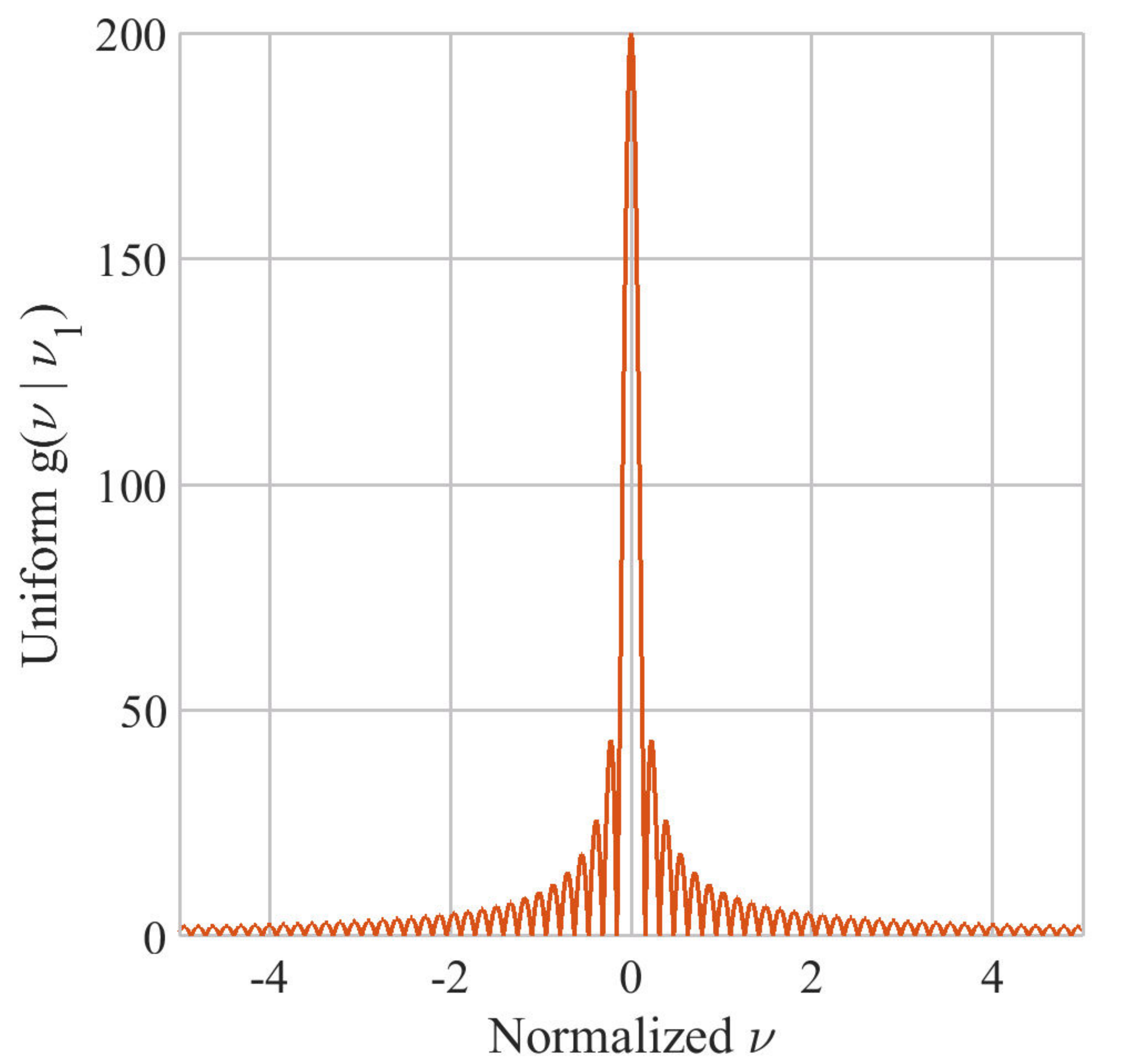}
    \vspace{-0.3cm}
	\caption{\scriptsize{Sum of exponents $|g(\nu|\nu_l)|$ for $P=200$, $\tau=1$sec and $\nu_l=0$.}\vspace{-5pt}}
	\label{fig:sum_exp}
\vspace{-0.5cm}
\end{figure}
We now turn to the radar receiver design and describe how a delay-Doppler map can be recovered from only $N_b$ transmitted narrow bands. The radar receiver first filters the transmitted bands supported on $\mathcal{F}_R$ given by (\ref{eq:omega}) and computes the Fourier coefficients of the received signal.

Consider the Fourier series representation of the aligned frames $r_{R_X}^p(t+p \tau)$, with $r_{R_X}^p(t)$ defined in (\ref{eq:one_frame}):
\begin{eqnarray}
\label{eq:fourier_coeff}
c_p[k] & = & \int_0^{\tau} r_{R_X}^p(t+p \tau) e^{-j 2\pi k t/\tau} \mathrm{d}t \nonumber \\
&=& \frac{1}{\tau} H[k] \sum_{l=0}^{L-1} \alpha_l e^{-j 2 \pi k \tau_l / \tau} e^{-j \nu_l p \tau},
\end{eqnarray}
for $k \in \kappa$, where $\kappa= \left\{k = \left. \left\lfloor \frac{f}{f_{\text{Nyq}}}N \right\rfloor \right| \, f \in \mathcal{F}_R \right\}$. From (\ref{eq:fourier_coeff}), we see that the unknown parameters $\{ \alpha_l, \tau_l, \nu_l \}_{l=0}^{L-1}$ are embodied in the Fourier coefficients $c_p[k]$. The goal is then to recover these parameters from $c_p[k]$ for $k \in \kappa$ and $0 \leq p \leq P-1$.

To that end, we adopt the Doppler focusing approach from \cite{barilan2014focusing}. Consider the DFT of the coefficients $c_p[k]$ in the slow time domain:
\begin{eqnarray}
\label{eq:focused_coeff}
\tilde{\Psi}_{\nu}[k] &=&\sum_{p=0}^{P-1} c_p[k] e^{j \nu p \tau}  \\
&=& \frac{1}{\tau} H[k] \sum_{l=0}^{L-1} \alpha_l e^{-j 2 \pi k \tau_l / \tau} \sum_{p=0}^{P-1} e^{j (\nu-\nu_l) p \tau}. \nonumber
\end{eqnarray}
The key to Doppler focusing follows from the approximation:
\begin{equation}
g(\nu|\nu_l) = \sum_{p=0}^{P-1} e^{j (\nu-\nu_l) p \tau} \approx \left\{ \begin{array}{ll} 
P & |\nu -\nu_l| < \pi /P \tau \\
0 & |\nu -\nu_l| \geq \pi /P \tau,
\end{array} \right.
\end{equation}
as illustrated in Fig.~\ref{fig:sum_exp}. Denote the normalized focused measurements $\Psi_{\nu}[k]$ so that
\begin{equation} \label{eq:focused_coeff_norm}
\Psi_{\nu}[k] = \frac{\tau}{PH[k]} \tilde{\Psi}_{\nu}[k].
\end{equation}

As in traditional pulse Doppler radar, suppose we limit ourselves to the Nyquist grid so that $\tau_l/\tau=r_l/N$, where $r_l$ is an integer satisfying $0 \leq r_l \leq N-1$. Then, (\ref{eq:focused_coeff_norm}) can be approximately written in vector form as
\begin{equation}
\label{eq:doppler_foc}
\mathbf{\Psi}_{\nu} =  \mathbf{F}_{\kappa} \mathbf{x}_\nu,
\end{equation}
where $\mathbf{\Psi}_{\nu} = \left[ \Psi_{\nu}[k_0] \dots \Psi_{\nu}[k_{K-1}] \right], k_i \in \kappa$ for $0 \leq i \leq K-1$, $\mathbf{F}_{\kappa}$ is composed of the $K$ rows of the $N \times N$ Fourier matrix indexed by $\kappa$ and $\mathbf{x}_{\nu}$ is a $L$-sparse vector that contains the values $\alpha_l$ at the indices $r_l$ for the Doppler frequencies $\nu_l$ in the ``focus zone", that is $|\nu -\nu_l| < \pi / P \tau$.
It is convenient to write (\ref{eq:doppler_foc}) in matrix form, by vertically concatenating the vectors $\mathbf{\Psi}_{\nu}$, for $\nu$ on the Nyquist grid, namely $\nu=-\frac{1}{2\tau}+\frac{1}{P\tau}$, into the $K \times P$ matrix $\bf \Psi$, as
\begin{equation}
\mathbf{\Psi} =  \mathbf{F}_{\kappa} \mathbf{X}.
\end{equation}
The $P$ equations (\ref{eq:doppler_foc}) can be solved simultaneously using Algorithm \ref{algo:focusing}, where in each iteration, the maximal projection of the observation vectors onto the measurement matrix are retained. The algorithm termination criterion follows from the generalized likelihood ratio test (GLRT) based framework presented in \cite{scharf1994matched}. For each iteration, the alternative and null hypotheses in the GLRT problem define the presence or absence of a candidate target, respectively. In the Algorithm, $Q{\chi_2^2(\rho)}$ denotes the right-tail probability of the chi-square distribution function with $2$ degrees of freedom, $\Lambda^C$ is the complementary set of $\Lambda$ and 
\begin{equation}
\rho=\frac{P_T}{\sigma^2 |\mathcal{F}_R|}
\end{equation}
is the SNR with $\sigma^2$ the noise variance and $P_T$ defined in (\ref{eq:pt}). 

\begin{algorithm}[!t]
\caption{Cognitive Radar}\label{algo:focusing} 
		\begin{algorithmic}[1]
		\qinput Observation vectors $c_p[k]$, for all $0 \leq p \leq P-1$ and $k \in \kappa$, probability of false alarm $P_{\text{fa}}$, noise variance $\sigma^2$, transmitted power $P_T$, total transmitted bandwidth $|\mathcal{F}_R|$
		\qoutput Estimated target parameters $\{ \hat{\alpha}_l, \hat{\tau}_l, \hat{\nu}_l \}_{l=0}^{L-1}$
		\State Create $\mathbf{\Psi}$ from $c_p[k]$ using fast Fourier transform (FFT) (\ref{eq:focused_coeff}), for $k \in \kappa$ and $\nu=-1/(2\tau)+p/(P\tau)$ for $0 \leq p \leq P-1$
        \State Compute detection threshold
		$$
		\rho = \frac{P_T}{\sigma^2 |\mathcal{F}_R|}, \quad \gamma = Q^{-1}_{\chi_2^2(\rho)}(1-\sqrt[N]{1-P_{\text{fa}}})
		$$
		\State Initialization: residual $\mathbf{R}_0=\mathbf{\Psi}$, index set $\Lambda_0=\emptyset$, $t=1$
		\State Project residual onto measurement matrix:
		$$
		\mathbf{\Phi} =\mathbf{F}_{\kappa}^H \mathbf{R}_{t-1}
		$$
		\State Find the two indices $\lambda_t = [\lambda_t(1) \quad \lambda_t(2)]$ such that
		$$
		[\lambda_t(1) \quad \lambda_t(2)] = \text{ arg max}_{i,j} \left| \mathbf{\Phi}_{i,j} \right|
		$$
		\State Compute the test statistic
		$$
		\Gamma=\frac{(\mathbf{F}_{\kappa})_{\lambda_t(1)}((\mathbf{R}_{t-1})_{\lambda_t(2)})^H((\mathbf{F}_{\kappa})_{\lambda_t(1)})^H(\mathbf{R}_{t-1})_{\lambda_t(2)}}{\sigma^2}
		$$
		where $(\mathbf{M})_i$ denotes the $i$th column of $\bf M$
		\State If $\Gamma > \gamma$ continue, otherwise go to step 12
		\State Augment index set $\Lambda_t = \Lambda_t  \bigcup \{ \lambda_t \}$
		\State Find the new signal estimate
		$$
		\mathbf{\hat{X}}_{t|\Lambda_t} = (\mathbf{F}_{\kappa})_{\Lambda_t}^{\dagger} \mathbf{\Psi}, \quad 			\mathbf{\hat{X}}_{t|\Lambda_t^C} =\mathbf{0}
		$$
		\State Compute new residual
		$$
		\mathbf{R}_t= \mathbf{\Psi}-(\mathbf{F}_{\kappa})_{\Lambda_t}\mathbf{\hat{X}}
		$$
		\State Increment $t$ and return to step 4
		\State Estimated support set $\hat{\Lambda}= \Lambda_t$
		\State $\hat{\tau}_l=\frac{\tau}{N} \hat{\Lambda}(l,1)$, $\hat{\nu}_l= \frac{1}{P\tau} \hat{\Lambda}(l,2)$, $\hat{\alpha}_l=\hat{\mathbf{X}}_{\hat{\Lambda}(l,1), \hat{\Lambda}(l,2)}$
	\end{algorithmic}
\end{algorithm}

The following theorem from \cite{barilan2014focusing} derives a necessary condition on the minimal number of samples $K$ and pulses $P$ for perfect recovery in a noiseless environment.
\begin{theorem} \cite{barilan2014focusing} \label{th:min}
The minimal number of samples required for perfect recovery of $\{\alpha_l, \tau_l, \nu_l\}$ with $L$ targets in a noiseless environment is $4L^2$, with $K \geq 2L$ and $P \geq 2L$.
\end{theorem}
Theorem \ref{th:min} translates into requirements on the total bandiwdth of the transmitted bands, such that
\begin{equation} \label{eq:min_bandwidth}
B_{\text{tot}}=N \sum_{i=1}^{N_b} \left\lceil \frac{B_r^i}{B_h} \right\rceil
\geq 2L.
\end{equation}
The multiband design strategy, besides allowing a dynamic form of the transmitted signal spectrum over only a small portion of the whole bandwidth to enable spectrum sharing, has two additional advantages. First, our CS reconstruction achieves the same resolution as traditional Nyquist processing over a significantly smaller bandwidth. Second, since we only use narrow bands to transmit, the whole power is concentrated in them. Therefore, the SNR in the sampled bands is improved.

Our resulting spectrum sharing SpeCX framework is summarized in Algorithm \ref{algo:sharing}.

\begin{algorithm}[!t]
\caption{Spectral Coexistence via Xampling (SpeCX)}\label{algo:sharing} 
	\begin{algorithmic}[1]
	\qinput Comm signal $x_C(t)$
	\qoutput Estimated target parameters $\{ \hat{\alpha}_l, \hat{\tau}_l, \hat{\nu}_l \}_{l=0}^{L-1}$
	\State Initialization: perform spectrum sensing at the receiver on $x_C(t)$ using Algorithm \ref{algo:comm} with $S_R=\emptyset$
	\State Choose the least noisy subbands for the radar transmit spectrum with respect to detected $\mathcal{F}_C$ using Algorithm \ref{algo:sel}
	\State Communicate the transmitted radar signal support $\mathcal{F}_R$ to the comm and radar receivers
    \State Perform target delay and Doppler estimation using Algorithm \ref{algo:focusing}
	\State Perform spectrum sensing at the comm receiver on $x(t)=x_C(t)+x_R(t)$ using Algorithm \ref{algo:comm}
	\State If $\mathcal{F}_C$ changes, the radar transmitter goes back to step 2
	\end{algorithmic}
\end{algorithm}
\vspace{-6pt} 
\section{Software and Hardware Experiments}
\label{sec:exp}

In this section, we present software and hardware simulations to illustrate our SpeCX framework. Software experiments illustrate the comm band detection performance of the CRo and the target detection by the CRr. Hardware simulations demonstrate the practical implementation of the SpeCX system.
\vspace{-.4cm}
\subsection{Software Simulations}
\label{subsec:swsim}

\begin{figure}[!t]
	\centering
    \vspace{-.3cm}
	\includegraphics[width=0.42\textwidth]{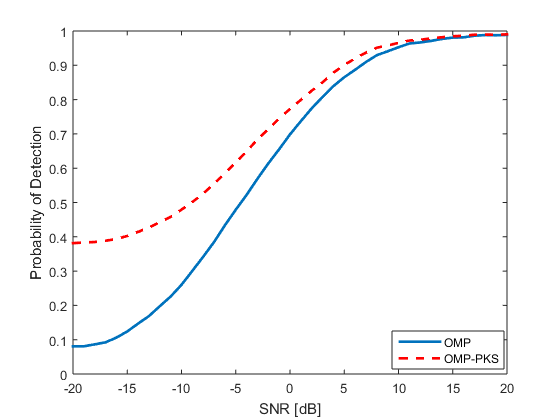}
    \vspace{-.3cm}
	\caption{Detection performance of the comm receiver in the presence of a radar signal with known support.}
    \vspace{-.5cm}
	\label{fig:sim7}
\end{figure}
To test the radio receiver, we consider a comm signal composed of $N_{\text{sig}}=2$ transmissions and a radar signal composed of $N_b=4$ bands with known support. The Nyquist rate is $f_{\text{Nyq}}=10\,\text{GHz}$. Each comm transmission has a two-sided bandwidth $B_c^i=50\,\text{MHz}$ and is modulated with a carrier $f_c^i$ drawn uniformly at random between $\pm f_{\text{Nyq}}/2=\pm5 \,\text{GHz}$. The CRo receiver is composed of $M=25$ analog channels, each sampling at rate $f_s=154\,\text{MHz}$ and with $K=91$ samples per channel. This leads to $N=195$ spectral bands. Figure \ref{fig:sim7} shows the performance of the detector for different values of the SNR, where the probability of detection is computed as the ratio of the correctly detected support. It can be seen that OMP-PKS, that exploits the knowledge of the radar signal's support, outperforms traditional OMP, as expected.
\begin{figure}[!t]
	\centering
    \vspace{-.2cm}
	\includegraphics[width=0.42\textwidth]{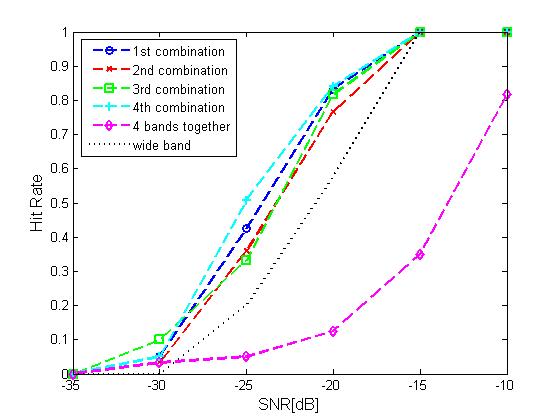}
    \vspace{-.3cm}
	\caption{Multiband versus wideband radar.}
    \vspace{-.5cm}
	\label{fig:soft_radar}
\end{figure}

For the radar receiver, we consider a transmission with $N_b=4$ spectral bands, each of bandwidth $81\,\text{KHz}$, yielding a total bandwidth of $324\,\text{KHz}$. For comparison, we simulate a wideband Nyquist pulse Doppler radar transmitting over a bandwidth $B_h=1.62\,\text{MHz}$. The cognitive radar thus transmits over only $20\%$ of the wideband. We consider $P=100$ pulses with PRI $\tau=10 \mu \, \text{sec}$. We use a hit-or-miss criterion as performance metric. A ``hit" is defined as a delay-Doppler estimate circumscribed by an ellipse around the true target position in the time-frequency plane. We used ellipses with axes equivalent to $\pm 3$ times the time and frequency Nyquist bins, defined as $1/B_h$ and $1/P\tau$, respectively. Figure~\ref{fig:soft_radar} shows the hit rate performance of our recovery method for different combinations of the transmitted spectral bands, which outperforms traditional wideband radar transmission and processing. Obviously, transmitting over adjacent bands yields poor results.
\vspace{-18pt} 
\subsection{Hardware Demo}
\label{subsec:hwdemo}
\begin{figure*}[!t]
	\centering
		\includegraphics[width=0.9\textwidth]{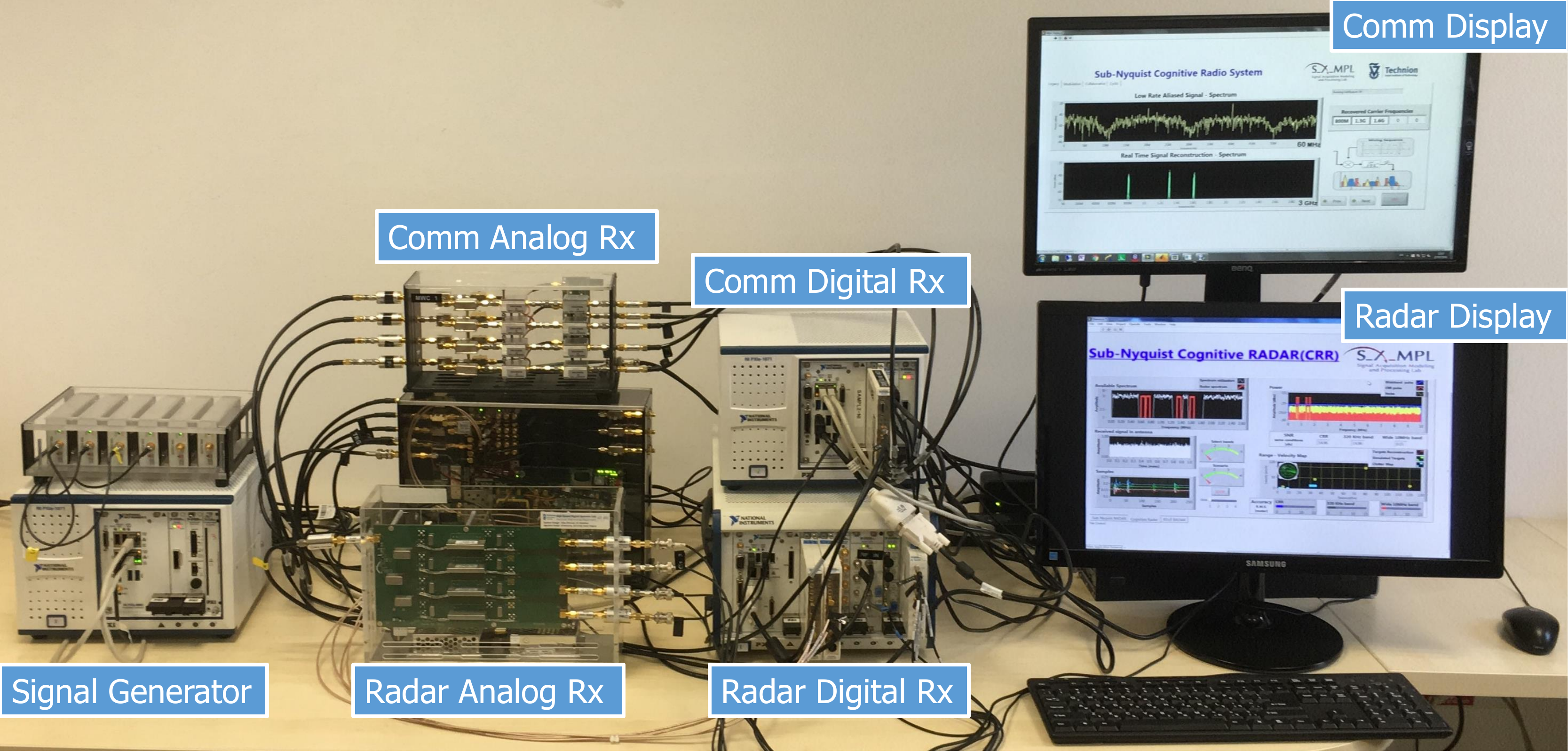}
		\caption{\scriptsize{Shared spectrum prototype. The system is composed of a signal generator, a CRo comm analog receiver including the MWC analog front-end board and the FPGA mixing sequences generator, a comm digital receiver, a CRr analog and receiver.}} 
		\label{fig:proto}
\end{figure*}
The SpeCX prototype, shown in Fig.~\ref{fig:proto}, is composed of a CRo receiver and a CRr transceiver. The CRo hardware realizes the system shown in Fig.~\ref{fig:HighLevel}. At the heart of the system lies the proprietary developed MWC board \cite{mishali2011hardware} that implements the sub-Nyquist analog front-end receiver. The card first splits the wideband signal into $M=4$ hardware channels, with an expansion factor of $q=5$, yielding $Mq=20$ virtual channels after digital expansion. In each channel, the signal is then mixed with a periodic sequence $p_i(t)$, generated on a dedicated FPGA, with $f_p=20\, \text{MHz}$. The sequences are chosen as truncated versions of Gold Codes \cite{gold1967optimal}, commonly used in telecommunication (CDMA) and satellite navigation (GPS). These were heuristically found to give good detection results in the MWC system \cite{mishali2009expected}, primarily due to small bounded cross-correlations within a set. This is useful when multiple devices are broadcasting in the same frequency range. 

Next, the modulated signal passes through an analog anti-aliasing LPF. Specifically, a Chebyshev LPF of 7th order with a cut-off frequency ($-3\,\text{dB}$) of $50\,\text{MHz}$ was chosen for the implementation. Finally, the low rate analog signal is sampled by a National Instruments\textsuperscript{\textcopyright} ADC operating at $f_s=(q+1)f_p=120\,\text{MHz}$ (with intended oversampling), leading to a total sampling rate of $480\,\text{MHz}$. The digital receiver is implemented on a National Instruments\textsuperscript{\textcopyright} PXIe-1065 computer with DC coupled ADC. Since the digital processing is performed at the low rate $120 \, \text{MHz}$, very low computational load is required in order to achieve real time recovery. MATLAB\textsuperscript{\textregistered}and LabVIEW\textsuperscript{\textregistered} platforms are used for the various digital recovery operations. 
\begin{figure}[!t]
	\centering
		\includegraphics[width=0.85\columnwidth]{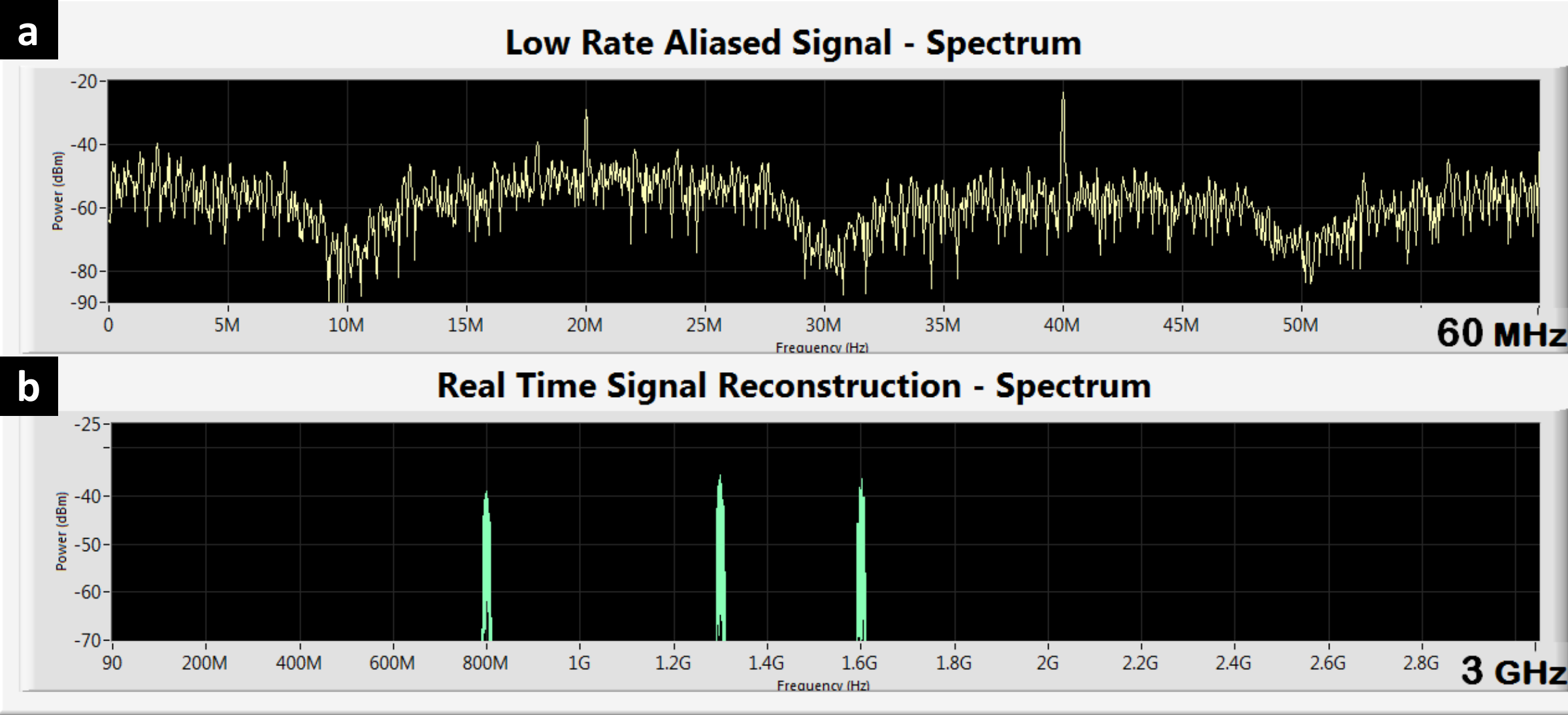}
        \vspace{-.2cm}
		\caption{\scriptsize{SpeCX comm system display showing (a) low rate samples acquired from one MWC channel at rate $ 120\,\text{MHz}$, and (b) digital reconstruction of the entire spectrum from sub-Nyquist samples.}}
		\label{fig:LegacySim}
\end{figure}

The prototype is fed with RF signals composed of up to $N_{\text{sig}}=5$ real comm transmissions, namely $K=10$ spectral bands with total bandwidth occupancy of up to $200\,\text{MHz}$ and varying support, with Nyquist rate of $6\,\text{GHz}$. 
Specifically, to test the system's support recovery capabilities, an RF input is generated using vector signal generators (VSG), each producing a modulated data channel with individual bandwidth of up to $20\,\text{MHz}$, and carrier frequencies ranging from $250\,\text{MHz}$ up to $3.1\,\text{GHz}$. The input transmissions then go through an RF combiner, resulting in a dynamic multiband input signal, that enables fast carrier switching for each of the bands. This input is specially designed to allow testing the system's ability to rapidly sense the input spectrum and adapt to changes, as required by modern CRo and shared spectrum standards, e.g. in the SSPARC program. The system's effective sampling rate, equal to $480\,\text{MHz}$, is only $8\%$ of the Nyquist rate and 2.4 times the Landau rate. This rate constitutes a relatively small oversampling factor of $20\%$ with respect to the theoretical lower sampling bound. The main advantage of the Xampling framework, demonstrated here, is that the sensing is performed in real-time from sub-Nyquist samples for the entire spectral range, which results in substantial savings in both computational and memory complexity. 

Support recovery is digitally performed on the low rate samples. The prototype successfully recovers the support of the comm transmitted bands, as demonstrated in Fig.~\ref{fig:LegacySim}. Once the support is recovered, the signal itself can be reconstructed from the sub-Nyquist samples in real-time. We note that the reconstruction does not require interpolation to the Nyquist rate and the active transmissions are recovered at the low rate of $20\,\text{MHz}$, corresponding to the bandwidth of the slices $\mathbf{z}(f)$. 

By combining both spectrum sensing and signal reconstruction, the MWC prototype serves as two separate comm devices. The first is a state-of-the-art CRo that performs real time spectrum sensing at sub-Nyquist rates, and the second is a unique receiver able to decode multiple data transmissions simultaneously, regardless of their carrier frequencies, while adapting to spectral changes in real time.
\begin{figure}[!t]
	\centering
	\includegraphics[width=0.45\textwidth]{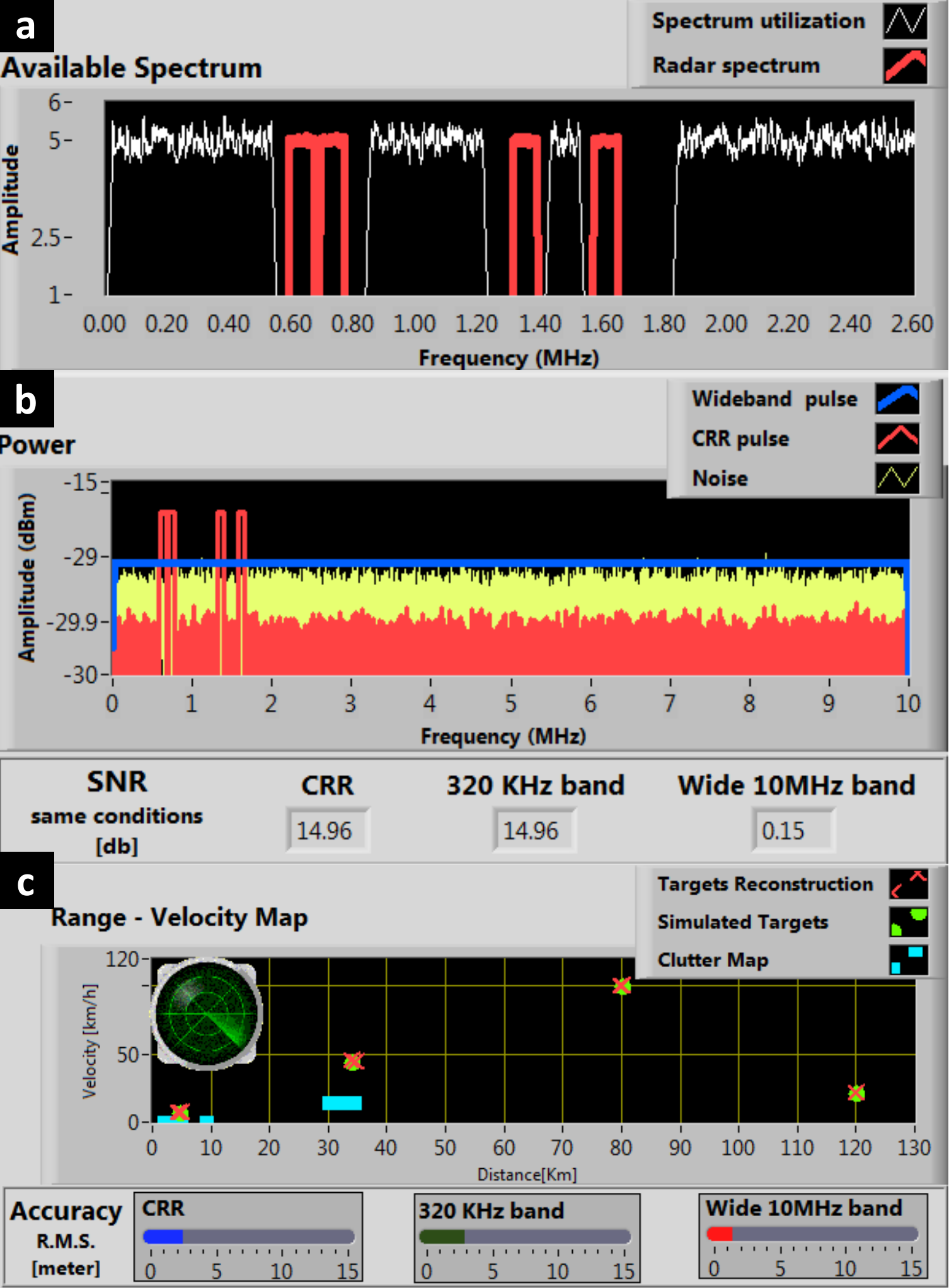}
    \vspace{-0.2cm}
	\caption{\scriptsize{SpeCX radar display showing (a) coexisting comm and CRr (b) CRr spectrum compared with the full-band radar spectrum, and (c) range-velocity display of detected and true locations of the targets.}}
	\label{fig:hardware}
\vspace{-0.5cm}
\end{figure}

The CRr system \cite{barilan2014focusing,baransky2014prototype, cohen2016towards} includes a custom made sub-Nyquist radar receiver board composed of $N_b=4$ parallel channels which sample distinct $N_b=4$ bands of the radar signal spectral content. In the $i$th channel, the transmitted band with center frequency $f_r^i$ and bandwidth $B_r^i=80 \, \text{KHz}$ is filtered, demodulated to baseband and sampled at $250 \, \text{KHz}$ (with intentional oversampling). This way, four sets of consecutive Fourier coefficients are acquired. More details on the hardware design can be found in \cite{baransky2014prototype}. After sampling, the spectrum of each channel output is computed via FFT and the 320 Fourier coefficients are used for digital recovery of the delay-Doppler map \cite{barilan2014focusing}.
The prototype simulates transmission of $P=50$ pulses towards $L=10$ targets. The CRr transmits over $N_b=4$ bands, selected according to the procedure presented in Section \ref{subsec:wf_opt}, after the spectrum sensing process has been completed by the comm receiver. We compare the target detection performance of our CRr with a traditional wideband radar with bandwidth $B_h=20\, \text{MHz}$. The CRr transmitted bandwidth is thus equal to $3.2\%$ of the wideband.

Figure~\ref{fig:hardware} shows windows from the GUI of our CRr system. Figure~\ref{fig:hardware}(a) illustrates the coexistence between the radar transmitted bands in red and the existing comm bands in white. The gain in power is demonstrated in Fig.~\ref{fig:hardware}(b); the wideband radar spectrum is shown in blue, our CRr in red and the noise in yellow in a logarithmic scale. The true and recovered range-velocity maps are shown in Fig.~\ref{fig:hardware}(c). All 10 targets are perfectly recovered and clutter, shown in blue, is discarded. Below the map, the range recovery accuracy is shown for 3 scenarios: from left to right, CRr in blue ($2.5$m), 4 adjacent bands with same bandwidth ($12.5$m) and wideband ($4$m). The poor resolution of the 4 adjacent bands scenario is due to its small aperture. Our CRr system with non-adjacent bands yields better resolution than the traditional wideband scenario.
\vspace{-6pt} 
\section{Summary}
\label{sec:summ}
Our SpeCX model proposes a comm and radar spectral coexistence approach through the well-established theory of Xampling. We demonstrated that the two networks can actively cooperate through handshaking of information on RF environment and optimize their performances. Unlike previous approaches, we presented a complete solution that shows signal recovery in both systems with the minimum of known information about the spectrum. We showed that the SpeCX is practically feasible through the development and real-time testing of our hardware prototype.

Some of the other elements of signal model that were not considered in this work include performance of the comm receiver when the radar signal is also contaminated with clutter and hostile jamming. Extensions to MIMO radar-comm spectrum sharing as described by \cite{li2016mimo} are also interesting, especially since that we recently demonstrated a hardware prototype of a cognitive sub-Nyquist MIMO radar \cite{mishra2016cognitive}. It would also be useful to incorporate additional optimization constraints in the radar waveform design.
\vspace{-12pt}
\bibliographystyle{IEEEtran}
\bibliography{refs}

\end{document}